\documentclass[proof]{WileyASNA-v1}

\articletype{Article Type}%
\received{xxx}
\revised{xxx}
\accepted{xxx}

\begin{document}

\title{Anisotropy in Magnetized Quark Matter in the Chiral Limit}

\author[1]{S.A. Ferraris*}
\author[2,3]{J.P. Carlomagno}
\author[2,3]{G.A. Contrera}
\author[1,3]{A.G. Grunfeld}

\authormark{S.A. Ferraris et al.}

\address[1]{\orgdiv{Physics Department}, \orgname{Comisión Nacional de Energía Atómica (C.N.E.A), Avenida del Libertador 8250}, \orgaddress{\state{ Ciudad Autónoma de Buenos Aires}, \country{Argentina}}}

\address[2]{\orgdiv{IFLP, CONICET Departamento de Física}, \orgname{Comisión Nacional de Energía Atómica (C.N.E.A)}, \orgaddress{\state{ Buenos Aires}, \country{Argentina}}}

\address[3]{\orgname{CONICET}, \orgaddress{Godoy Cruz 2290 (C1425FQB) CABA,\state{Ciudad Autónoma de Buenos Aires}, \country{Argentina} }}

\corres{*Av. Gral. Paz 1499, San Martin, Buenos Aires.  \email{sebsatianferraris@cnea.gob.ar}}

\abstract{
We investigate the behavior of cold quark matter under strong magnetic fields in the frame of a nonlocal NJL model in the chiral limit. Our analysis focuses on deconfinement, chiral symmetry restoration, and the anisotropy in pressure induced by the external magnetic field.

For $eB \lesssim 0.07 \, \text{GeV}^2$, the critical chemical potential remains largely insensitive to the magnetic field, whereas at higher field strengths, transitions to chirally restored phases occur at progressively lower chemical potentials. The parallel and perpendicular pressures, respect to the magnetic field, exhibit distinct behaviors, reflecting the anisotropic nature of the system. Oscillations in the quark number density, driven by the de Haas–van Alphen effect, reflect the quantized behavior of quarks in a magnetic field. Similarly, the magnetization displays oscillatory behavior, driven by the sequential filling of Landau levels. At lower external magnetic field strengths, contributions from orbital angular momentum and the population of higher Landau levels further modulate these oscillations.

These results provide deeper insights into the thermodynamic and magnetic properties of quark matter under strong magnetic fields, with implications for astrophysical studies.}


\maketitle

\section{Introduction}

Magnetized quark matter holds significant relevance in astrophysics, offering insights into the behavior of strongly interacting matter under extreme conditions, particularly in the environments of neutron stars. Magnetars, a subclass of neutron stars, possess magnetic fields up to $\sim 10^{15} \, \text{Gauss}$, which profoundly influence the thermodynamic and structural properties of matter. These extreme magnetic fields can lead to exotic phases, including color superconductivity, and induce anisotropies that substantially alter the equation of state (EOS). Such effects are crucial for understanding the maximum mass, radius, and tidal deformability of this kind of compact stars. 

Similarly, magnetized quark matter is essential in the study of heavy-ion collisions, where transient magnetic fields as strong as $\sim 10^{18} \, \text{Gauss}$ are generated. These fields introduce anisotropies in the quark-gluon plasma (QGP), affecting its thermodynamic properties, including pressure, sound speed, and particle production. The presence of a strong magnetic field ($eB$) breaks spatial rotational symmetry, leading to anisotropies in the energy-momentum tensor. This results in distinct pressures along and perpendicular to the magnetic field direction, which have profound implications for the EOS and the stability of magnetized matter.

In this work, we focus on the region of the QCD phase diagram most relevant to astrophysical applications, characterized by zero temperature and finite chemical potential. This region is of particular interest due to its relevance to the cores of neutron stars, where dense, strongly interacting quark matter may exist. Using the nonlocal Nambu–Jona-Lasinio (nlNJL) model in the chiral limit, we analyze the anisotropic thermodynamic properties of magnetized quark matter, such as pressure and magnetization. The nlNJL model is well-suited for this purpose, as it incorporates nonlocal interactions and reproduces essential QCD features, including chiral symmetry breaking and inverse magnetic catalysis (IMC).

Our study reveals the anisotropic behavior of pressure, which result from the system’s response to compression along different spatial directions under a strong magnetic field. Additionally, we investigate the de Haas–van Alphen effect, observed as oscillations in the quark number density and magnetization due to the sequential filling of Landau levels. These phenomena provide valuable insights into the quark matter behavior in a magnetized environment, offering potential implications for the structure of neutron stars and the evolution of the QGP in heavy-ion collisions. 

While astrophysical applications of our results remain beyond the scope of this study, they represent an important direction for future research.

This work is organized as follows: Sec.~2 introduces the theoretical formalism of the nlNJL model for magnetized quark matter in the chiral limit at zero temperature. Sec.~3 presents the numerical results, focusing on the anisotropies in pressure and magnetization, and Sec.~4 concludes with a summary of the main findings and their relevance to astrophysical and experimental contexts.

\section{Formalism}
In this work, we use the nonlocal chiral quark model framework described in Ref.~\cite{Ferraris:2021vzm}, where the full details of the calculations are provided. We focus on a system's grand canonical thermodynamic potential for two light flavours $u,d$, at zero temperature and finite quark chemical potential \( \mu \), under an external constant magnetic field in $z$-direction. In the mean-field approximation (MFA), it is given by

\begin{eqnarray}
\Omega_{\mu,B}^{\text{MFA}} &=& \frac{\bar{\sigma}^2}{2G} - 3 \sum_{f=u,d} \frac{|q_f B|}{2\pi} \int \frac{d^2p_{||}}{(2\pi)^2} \nonumber \\
&& \times \left[ \ln\left(p_{||}^2 + \left(M_{0,p_{||}}^{s_f,f}\right)^2\right) + \sum_{k=1}^\infty \ln \Delta_{k,p_{||}}^f \right],
\end{eqnarray}
where \( \Delta_{k,p_{||}}^f \), \(M_{k,p_{||}}^{\lambda,f}\) and \( g_{k,p_{||}}^{\lambda,f} \) are given by:
\begin{equation}
\Delta_{k,p_{||}}^f = \left(2k|q_f B| + p_{||}^2 + M_{k,p_{||}}^{+,f} M_{k,p_{||}}^{-,f}\right)^2 + p_{||}^2\left(M_{k,p_{||}}^{+,f} - M_{k,p_{||}}^{-,f}\right)^2,
\end{equation}
\begin{equation}
M_{k,p_{||}}^{\lambda,f} = \left(1-\delta_{k_\lambda,-1}\right)m_c + \bar{\sigma} g_{k,p_{||}}^{\lambda,f},
\end{equation}
\begin{equation}
\begin{split}
g_{k,p_{||}}^{\lambda,f}&=\frac{4\pi}{|q_{f}B|}\left(-1\right)^{k_{\lambda}}\int \frac{d^{2}p_{\bot}}{\left(2\pi\right)^{2}}\ g\left(p_{\bot}^{2} + p_{||}^{2}\right)\\
&\times exp\left(-p_{\bot}^{2}/|q_{f}B|\right)L_{k_{\lambda}}\left(2p_{\bot}^{2}/|q_{f}B|\right).
\label{funcg}
\end{split}
\end{equation}
We define \( p_\bot = (p_1, p_2) \), \( p_{||} = (p_3, p_4 - i\mu) \), \( k_\pm = k - 1/2 \pm s_f/2 \), with \( s_f = \text{sign}(q_f B) \) and \( L_m(x) \) as Laguerre polynomials.
It is important to note that we work in the chiral limit, which implies setting \( m_c = 0 \) in all the expressions.
The Gaussian form factor is given by:
\begin{equation}
g(p^2) = e^{-p^2 / \Lambda^2}.
\end{equation}
With this, the integrals in \( g_{k,p_{||}}^{\lambda,f} \) can be solved analytically, leading to:
\begin{equation}
M_{\bar{p},k}^{\lambda,f} = \bar{\sigma} \frac{\left(1-|q_f B|/\Lambda^2\right)^{k_\lambda}}{\left(1+|q_f B|/\Lambda^2\right)^{k_\lambda+1}} e^{-p_{||}^2/\Lambda^2}.
\end{equation}

The thermodynamic potential $\Omega_{\mu,B}^{\text{MFA}}$ is divergent and can be regularized as follows
\begin{equation}
\Omega_{\mu,B}^{\text{MFA,reg}} = \Omega_{\mu,B}^{\text{MFA}} - \Omega_{\mu,B}^{free} + \Omega_{\mu,B}^{free,reg},
\end{equation}
where \( \Omega_B^{free} \) is computed by setting \( \bar{\sigma} = 0 \) and the last term is taken from Ref. \cite{Menezes:2008qt}. Then, the regularized thermodynamic potential reads:

\begin{align}
\Omega_{\mu,B}^{\text{MFA,reg}} =\ & \frac{\bar{\sigma}^2 }{2G}+N_{c}\sum_{f=u,d} \frac{|q_{f}B|}{2\pi} \int{\frac{dp_{3} dp_{4}}{4\pi^{2}}}~ \mathcal{F}_{k}^{f}(p_{||})\nonumber\\
& -\frac{N_{c}}{2\pi^{2}}\sum_{f=u,d}\left(q_{f}B\right)^{2}~ \mathcal{G}_{k,p_{||}}^{f}\nonumber\\
& -\frac{N_{c}}{4\pi^{2}} \sum_{k,f=u,d} \theta\left(\mu-S_{kf}\right) \alpha_{k}|q_{f}B|~\mathcal{H}_{k,p_{||}}^{f}
\end{align}

were
\begin{align}
 \mathcal{F}_{k}^{f}(p_{||})=& ~ln\left( \frac{p_{||}^{2} }{p_{||}^{2} + \left(M_{0,p_{||}}^{\lambda,f}\right)^{2} } \right) +\sum_{k=1}^{\infty} {ln\left(\frac{\left(2k |q_{f} B| + p_{||}^{2} \right)^{2}}{\Delta_{k,p_{||}}^{f}} \right)},\nonumber\\
 \mathcal{G}_{k,p_{||}}^{f}=& ~\zeta^{'}\left(-1,x_{f}\right)  + \frac{x_{f}^{2}}{4} -\frac{1}{2}\left(x_{f}^{2} - x_{f}\right) ln \left(x_{f}\right)\nonumber,\\
 \mathcal{H}_{k,p_{||}}^{f}=& ~ \mu\sqrt{\mu^{2}-S_{kf}^{2}} - S_{kf}^{2} ln \left(\frac{\mu + \sqrt{\mu^{2}-S_{kf}^{2}} }{S_{kf}}\right).\nonumber
\end{align}
Here we have used the definitions $x_{f}=m_{c}^{2}/(2|q_{f}B|)$, $S^2_{k,f} = 2k |q_{f} B|$,
$\alpha_{k}=2-\delta_{k,0}$, $\epsilon_{kp}^{f}=\left( (2k|q_{f}B|) +p^{2} \right)^{1/2}$ and
$\zeta^{'}(-1,x_{f})=d\zeta(z,x_{f})/dz|_{z=-1}$, where $\zeta(z,x_{f})$ is
the Hurwitz zeta function.

Finally, the mean field values \( \bar{\sigma} \) are obtained by solving the gap equation:
\begin{equation}
\frac{\partial \Omega_{\mu,B}^{\text{MFA,reg}}}{\partial \bar{\sigma}} = 0.
\end{equation}
This provides the thermodynamic properties of the system under an external magnetic field at finite chemical potential.

Next, we provide the explicit expressions for key thermodynamic quantities.  
The parallel and perpendicular pressures, \( p_{||} \) and \( p_\perp \), are defined as (see Refs. \cite{Ferrer:2010wz,Strickland:2012vu}):  
\begin{align}
p_{||} &= -\Omega_{\mu,B}^{\rm MFA,reg},\\
p_{\perp} &= -\Omega_{\mu,B}^{\rm MFA,reg} - eB~\mathcal{M}_q, 
\end{align}
where $\mathcal{M}_q$ stand for the magnetization and is defined as
\begin{equation}
 \mathcal{M}_{q}= - \left(\frac{\partial~\Omega_{\mu,B}^{\rm MFA,reg}}{\partial (eB)} \right).    
\end{equation}

\subsection{Chiral symmetry restored phase}

We seek to investigate the behavior of quark matter under intense magnetic fields in the chiral symmetry restored phase, which corresponds to the most dynamic and complex region of the QCD phase diagram. This phase is of particular interest due to the distinct thermodynamic and structural properties it exhibits as a result of the applied magnetic field. In this regime, especially in the chiral limit where the mean field \( \bar{\sigma} \) vanishes, the analytical expressions for the thermodynamic quantities become significantly simplified.

In this regime, the quark number density, defined as  
\begin{equation}
\rho_{q} = -\frac{\partial \Omega_{\mu,B}^{\rm MFA,reg}}{\partial \mu},
\end{equation}  
exhibits distinctive features associated with transitions linked to the \textit{de Haas–van Alphen} effect, which arise from the sequential filling of Landau levels.

For the lowest Landau level (LLL), the magnetization is given by  
\begin{align}
\mathcal{M}_{q,k=0} &= \frac{N_{c}}{4\pi^{2}} \sum_{f=u,d} |q_{f}| \mu^{3},
\end{align}  
where \( N_c \) is the number of color charges, \( q_f \) is the electric charge of the quark flavor \( f \), and \( \mu \) is the quark chemical potential.

For higher Landau levels (\( k \neq 0 \)), the magnetization takes the form  
\begin{align}
 \mathcal{M}_{k\neq0} &= \frac{N_{c}}{2\pi^{2}}\sum_{f=u,d} 2q_{f}^{2}B~\zeta^{'}(-1,0) \nonumber \\
&+\frac{N_{c}}{4\pi^{2}}\sum_{k,f}^{k_{max}}\theta(\mu-S_{k,f})\alpha_{k}|q_{f}| \times \nonumber \\ & \left[\mu\sqrt{\mu^{2}-S^2_{k,f}}-2 S^2_{k,f}~ln\left(\frac{\mu+\sqrt{\mu^{2}-S^2_{k,f}}}{S_{k,f} } \right) \right]  
\end{align} 
where the detailed expression reflects the contributions from the sequential filling of Landau levels.

\section{Results}

The model parameters are $\Lambda = 760.142$ MeV and $G \Lambda^2 = 20.605$ considered in Ref.~\cite{Ferraris:2021vzm}. These parameters are selected to obtain \( f_{\pi, ch} = 90 \, \text{MeV} \) in the chiral limit and a phenomenological quark condensate. Here we considered \( -\langle \bar{q}q \rangle^{1/3} = 240 \, \text{MeV} \).

\begin{figure}[htp]
 \begin{center}
\begin{tabular}{ c }
 \vspace{-0.8cm}
 \includegraphics[scale=0.30]{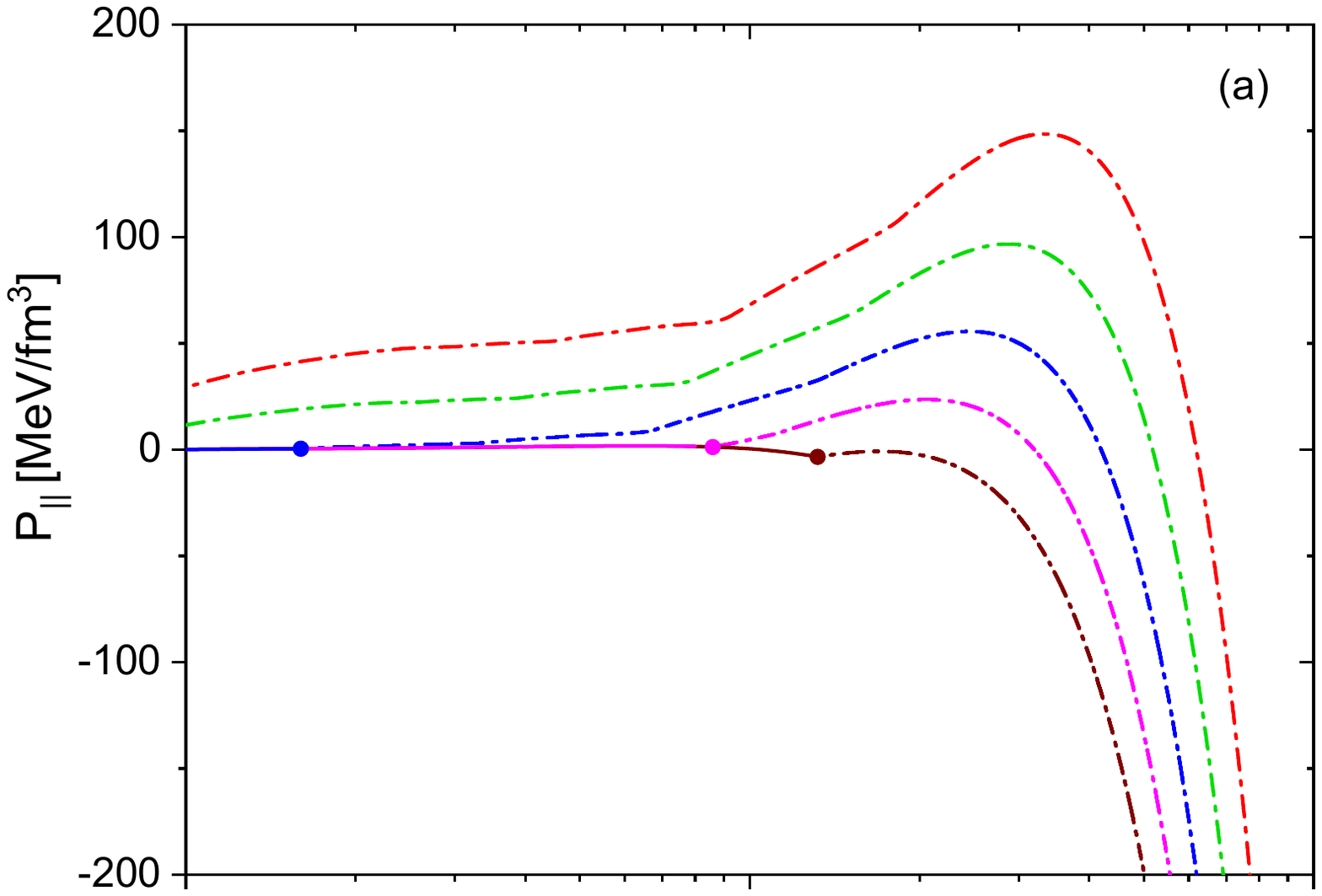}\\ 
 \includegraphics[scale=0.30]{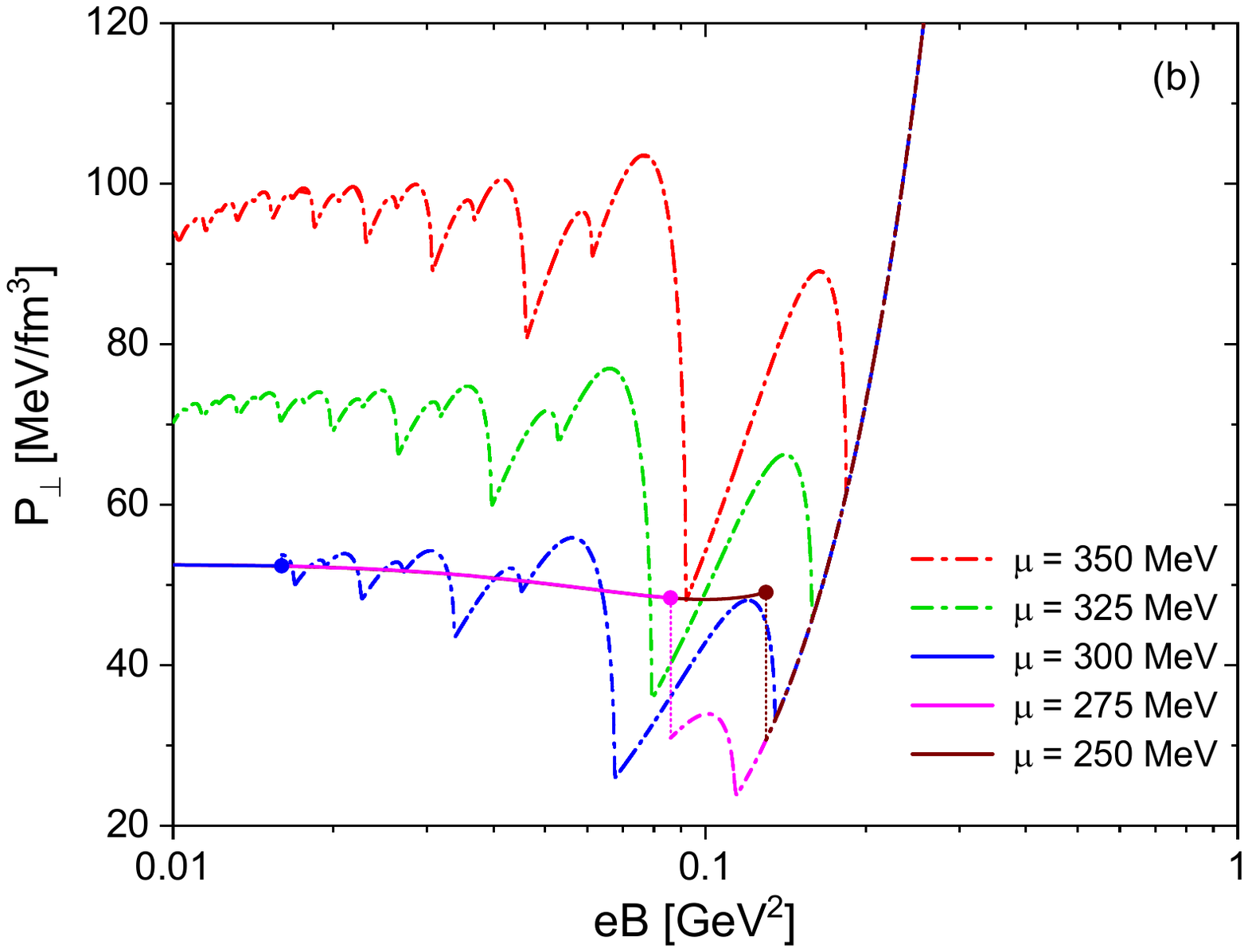}    \end{tabular}
\end{center}
\caption{Panel (a) shows \( P_{||} \) and panel (b) shows \( P_{\perp} \) as functions of the external magnetic field \( eB \). Solid lines correspond to the chiral symmetry broken phase, while dash-dotted lines represent the chiral symmetry restored phase. The dots indicate the phase transition from the chiral symmetry broken phase to the restored one. Different values of the chemical potential are considered for both panels.}
\label{figura1}
\end{figure}

As mentioned in the introduction, we aim to analyze the anisotropy in the pressure induced by the external strong constant magnetic field. In Fig.\ref{figura1} (a) and (b), we present the behavior of both \( p_{||} \) and \( p_{\perp} \) respectively, as functions of \( eB \) for different fixed values of the chemical potential.
To ensure that the pressure vanishes in the vacuum, i.e., at \( T = \mu = B = 0 \), we have subtracted a constant from the thermodynamic potential, as given by \( \Omega_{\mu,B}^{\rm MFA,reg} - \Omega_{\rm vac} \). However, from Fig.\ref{figura1}, we see that the normalized pressure \( p_{||} = -(\Omega_{\mu,B}^{\rm MFA,reg} - \Omega_{\rm vac}) \) becomes negative at a given magnetic field. Some authors \cite{Menezes:2015fla}, \cite{Chaudhuri:2022oru}, \cite{Goswami:2023eol}, avoid this behavior by subtracting \( \Omega_{\rm vac}(0,0,eB) \), ensuring that the pressure in the vacuum, immersed in a constant magnetic field, is zero.

\begin{figure}[htb]
\centering 
\includegraphics[width=0.5\textwidth]{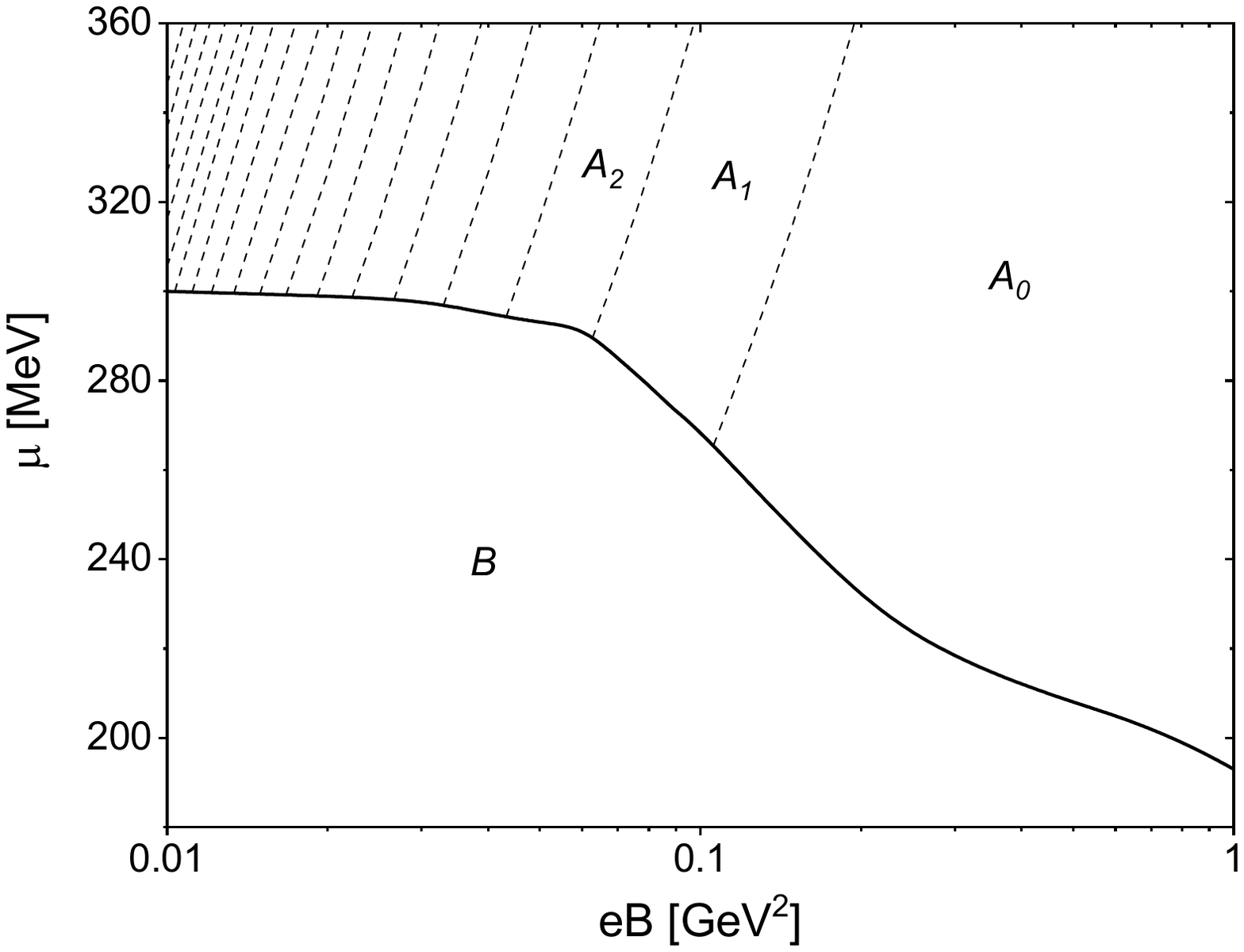}
\caption{Phase diagram in the \( \mu \)-\( eB \) plane. The phase labeled \( B \) represents the fully chiral broken phase, while the phases \( A_i \) (\( i = 0, 1, 2, \dots \)) correspond to regions where chiral symmetry is restored, with different numbers of populated Landau levels (LLs).}
\label{figura2}
\end{figure}

\begin{figure}[htb]
 \begin{center}
\begin{tabular}{ c }
\vspace{-0.8cm}
\includegraphics[scale=0.30]{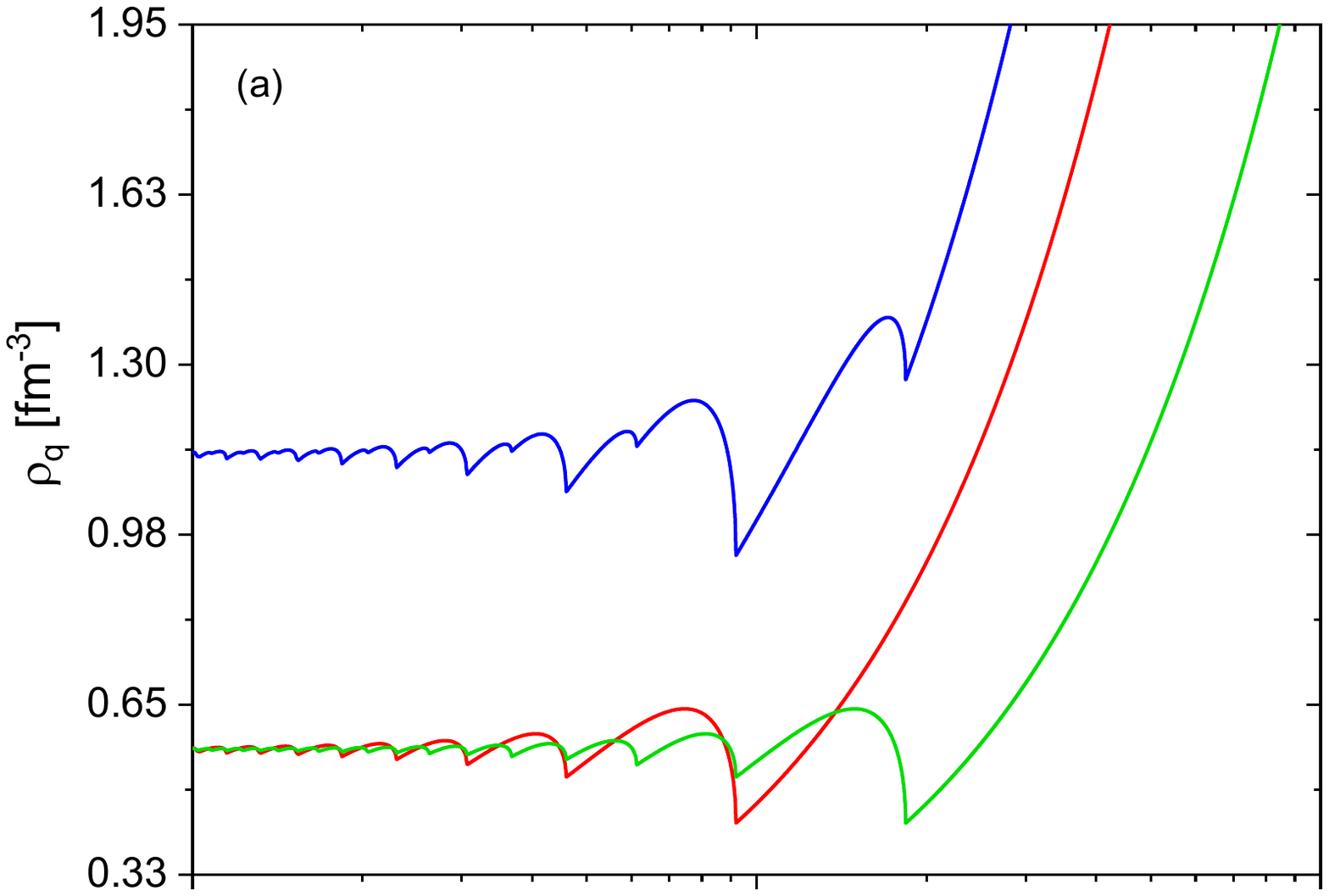}\\
\includegraphics[scale=0.30]{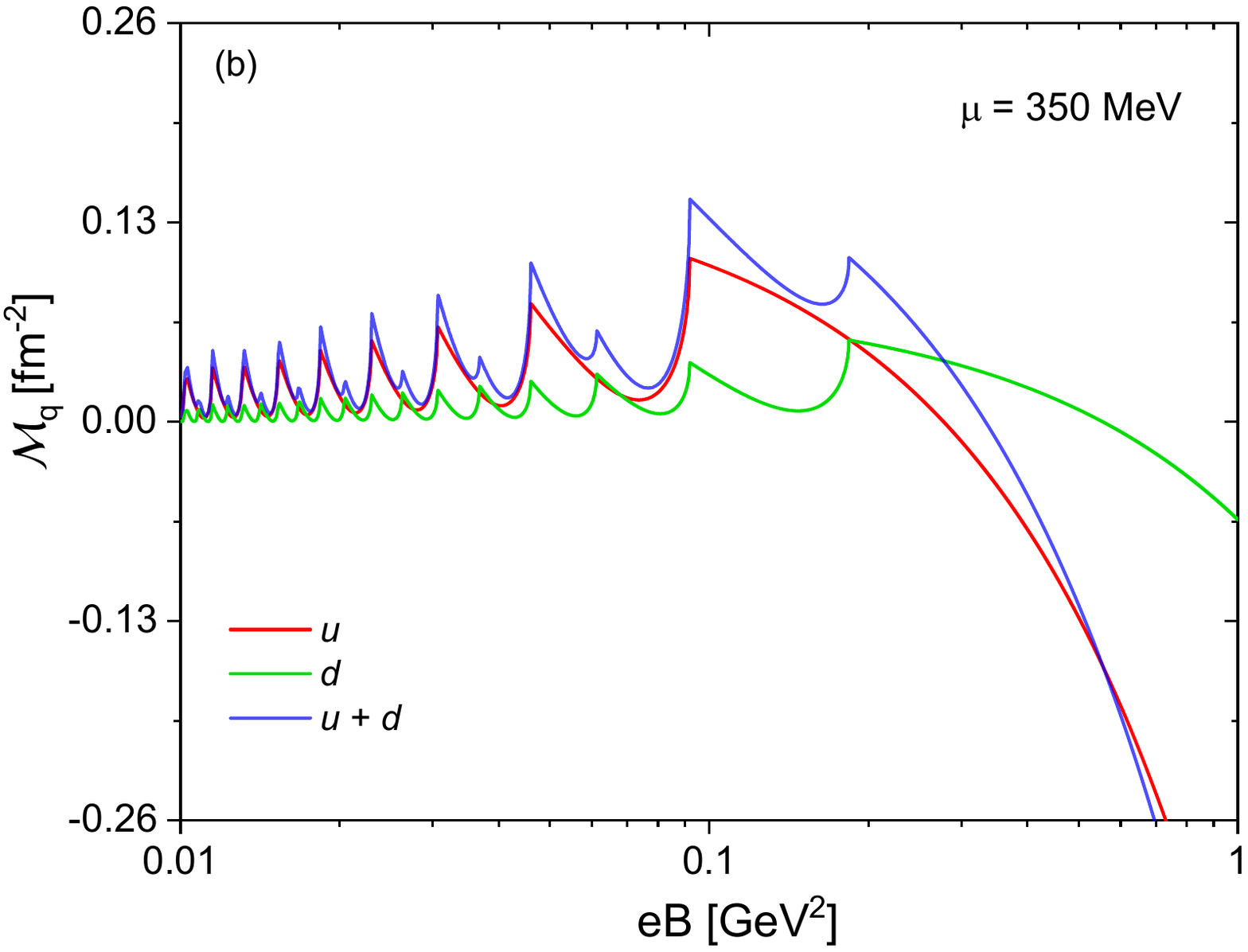}
\end{tabular}
\end{center}
\caption{ Panel (a) shows the quark number density \( \rho_q \) as a function of the external magnetic field \( eB \). Panel (b) shows the response of the system's magnetization \( \mathcal{M}_q \) to the applied magnetic field \( eB \), highlighting the evolution of magnetization as the field strength increases. In both panels, we have considered a fixed quark chemical potential \( \mu = 350 \, \text{MeV} \).
}
\label{figura3}
\end{figure}

In Fig. \ref{figura2} we present the phase diagram in the $\mu$–$eB$ plane constructed using the previously introduced parametrization. The phase labeled $B$ represents the fully chirally broken phase, whereas the phases denoted as $A_i$ ($i = 0,1,2,\ldots$) correspond to regions where chiral symmetry is restored. The $A_i$ phases signify distinct configurations with different numbers of populated Landau levels (LLs). 

We observe that for magnetic field values up to $eB \sim 0.07$ GeV$^2$, the critical chemical potential $\mu_c$ exhibits minimal sensitivity to changes in the magnetic field. Beyond this threshold, for higher $eB$ values, the transition to the $A_i$ phases occurs at progressively lower chemical potentials. This behavior contrasts with results obtained in local NJL models, where $\mu_c$ initially decreases with $eB$, reaches a minimum near $eB \sim 0.2$ GeV$^2$, and then increases for stronger magnetic fields (as reported in \cite{Ferraris:2021vzm}).

In Fig. \ref{figura3}(a), we illustrate the dependence of the quark number density on the strength of the external magnetic field. The appearance of successive peaks in the curve provides clear evidence of phase transitions associated with the de Haas–van Alphen effect. These oscillatory features arise from the sequential population of Landau levels as the magnetic field increases, reflecting the quantized nature of the quark dynamics in a magnetized environment.

In Fig. \ref{figura3} (b), we present the magnetization behavior as a function of the external magnetic field. We note that the magnetization becomes negative for the \textit{d}-quark flavor within the magnetic field range of approximately $0.55 - 1 , \text{GeV}^{2}$, and for the \textit{u}-quark flavor within $0.27 - 1 , \text{GeV}^{2}$. In these intervals, the system is confined to the lowest Landau level (LLL, $k=0$), where the angular momentum contribution arises solely from the particle’s spin. As a result, only one spin state is allowed for each quark flavor: \textit{u}-quarks with spin up and \textit{d}-quarks with spin down. The negative magnetization indicates that the quarks' spins are anti-aligned with the magnetic field, a characteristic behavior of diamagnetism.

For magnetic fields outside these intervals, particularly at lower field strengths, the magnetization tends to zero as the field approaches zero. This behavior suggests a tendency for the particle spins to align with the magnetic field, leading to a balance between spin-up and spin-down states. However, this interpretation is less accurate in these regimes, as the system no longer remains restricted to the LLL. In this context, the orbital angular momentum of the particles begins to contribute alongside the spin angular momentum. As the magnetic field further decreases, higher Landau levels become populated, and the contribution of orbital angular momentum becomes more significant.

Moreover, in these lower-field regimes, we observe distinct oscillations in the magnetization, which are attributed to the sequential filling of Landau levels.

\section{Conclusions}

In this work, we have investigated the impact of an external strong magnetic field on the thermodynamic and magnetic properties of quark matter within a nonlocal NJL model. The parameters considered, $\Lambda = 760.142$ MeV and $G \Lambda^2 = 20.605$, were chosen to reproduce a pion decay constant $f_{\pi, \text{ch}} = 90 \, \text{MeV}$ in the chiral limit and a phenomenological quark condensate of $-\langle \bar{q}q \rangle^{1/3} = 240 \, \text{MeV}$.

We analyzed the anisotropy in the pressures induced by the external magnetic field. From Fig.~\ref{figura1}, we observe that, while the parallel pressure $p_{||}$ decreases with the magnetic field, it becomes negative beyond a certain $eB$ value. This behavior arises due to the subtraction of the vacuum contribution to ensure zero pressure at $T = \mu = B = 0$, though alternative regularization schemes, as suggested in Refs.~\cite{Menezes:2015fla, Chaudhuri:2022oru,Goswami:2023eol}, can mitigate this issue.

The phase diagram in the $\mu$--$eB$ plane, shown in Fig.~\ref{figura2}, reveals that the critical chemical potential $\mu_c$ remains relatively insensitive to changes in the magnetic field for $eB \lesssim 0.07 \, \text{GeV}^2$. At higher field strengths, the transitions to the chirally restored $A_i$ phases occur at progressively lower $\mu_c$ values. This trend contrasts with results from local NJL models, where $\mu_c$ decreases with $eB$ up to $eB \sim 0.2 \, \text{GeV}^2$ before increasing to stronger fields, as reported in Ref.~\cite{Ferraris:2021vzm}.

Figure~\ref{figura3}(a) highlights the oscillatory behavior of the quark number density, which reflects the de Haas–van Alphen effect due to the sequential filling of Landau levels. This result underscores the quantized nature of quark dynamics in a magnetized environment. The magnetization, shown in Fig.~\ref{figura3}(b), further reveals a transition to diamagnetic behavior at specific magnetic field intervals where the system is confined to the lowest Landau level (LLL). Within these ranges, the spin alignment of \textit{u}-quarks (spin-up) and \textit{d}-quarks (spin-down) becomes anti-aligned with the magnetic field.

At lower magnetic field strengths, the magnetization tends toward zero as the field approaches zero, reflecting a balance between spin-up and spin-down states. In these regimes, contributions from the orbital angular momentum become significant, and oscillatory patterns emerge as higher Landau levels are populated.

In summary, this study provides a comprehensive analysis of the anisotropic pressures, phase transitions, and magnetic properties of quark matter under strong magnetic fields. Future work may extend this analysis to explore the implications of these findings in astrophysical and heavy-ion collision contexts.

\section*{acknowledgments}
This work was partially supported bym CONICET under Grant No. PIP 22-24 11220210100150CO, ANPCyT (Argentina) under Grant PICT20-01847 and PICT19-00792, and the National University of La Plata (Argentina), Project No. X824.

\bibliography{Template}%

\begin{thebibliography}{}

\bibitem [\protect \citeauthoryear {%
Chaudhuri%
, Ghosh%
, Roy%
\BCBL {}\ \BBA {} Sarkar%
}{%
Chaudhuri%
\ \protect \BOthers {.}}{%
{\protect \APACyear {2022}}%
}]{%
Chaudhuri:2022oru}
\APACinsertmetastar {%
Chaudhuri:2022oru}%
\begin{APACrefauthors}%
Chaudhuri, N.%
, Ghosh, S.%
, Roy, P.%
\BCBL {}\ \BBA {} Sarkar, S.%
\end{APACrefauthors}%
\unskip\
\newblock
\APACrefYearMonthDay{2022}{}{},
\newblock
\unskip
\newblock
\APACjournalVolNumPages{Phys. Rev. D}{106}{5}{056020}.
\newblock
\begin{APACrefDOI} \doi{10.1103/PhysRevD.106.056020} \end{APACrefDOI}
\PrintBackRefs{\CurrentBib}

\bibitem [\protect \citeauthoryear {%
Ferraris%
, Grunfeld%
\BCBL {}\ \BBA {} Scoccola%
}{%
Ferraris%
\ \protect \BOthers {.}}{%
{\protect \APACyear {2021}}%
}]{%
Ferraris:2021vzm}
\APACinsertmetastar {%
Ferraris:2021vzm}%
\begin{APACrefauthors}%
Ferraris, S\BPBI A.%
, Grunfeld, A\BPBI G.%
\BCBL {}\ \BBA {} Scoccola, N\BPBI N.%
\end{APACrefauthors}%
\unskip\
\newblock
\APACrefYearMonthDay{2021}{}{},
\newblock
\unskip
\newblock
\APACjournalVolNumPages{Astron. Nachr.}{342}{1-2}{469--472}.
\newblock
\begin{APACrefDOI} \doi{10.1002/asna.202113952} \end{APACrefDOI}
\PrintBackRefs{\CurrentBib}

\bibitem [\protect \citeauthoryear {%
Ferrer%
, de~la Incera%
, Keith%
, Portillo%
\BCBL {}\ \BBA {} Springsteen%
}{%
Ferrer%
\ \protect \BOthers {.}}{%
{\protect \APACyear {2010}}%
}]{%
Ferrer:2010wz}
\APACinsertmetastar {%
Ferrer:2010wz}%
\begin{APACrefauthors}%
Ferrer, E\BPBI J.%
, de~la Incera, V.%
, Keith, J\BPBI P.%
, Portillo, I.%
\BCBL {}\ \BBA {} Springsteen, P\BPBI L.%
\end{APACrefauthors}%
\unskip\
\newblock
\APACrefYearMonthDay{2010}{}{},
\newblock
\unskip
\newblock
\APACjournalVolNumPages{Phys. Rev. C}{82}{}{065802}.
\newblock
\begin{APACrefDOI} \doi{10.1103/PhysRevC.82.065802} \end{APACrefDOI}
\PrintBackRefs{\CurrentBib}

\bibitem [\protect \citeauthoryear {%
Goswami%
, Sahu%
, Dey%
, Sahoo%
\BCBL {}\ \BBA {} Stock%
}{%
Goswami%
\ \protect \BOthers {.}}{%
{\protect \APACyear {2024}}%
}]{%
Goswami:2023eol}
\APACinsertmetastar {%
Goswami:2023eol}%
\begin{APACrefauthors}%
Goswami, K.%
, Sahu, D.%
, Dey, J.%
, Sahoo, R.%
\BCBL {}\ \BBA {} Stock, R.%
\end{APACrefauthors}%
\unskip\
\newblock
\APACrefYearMonthDay{2024}{}{},
\newblock
\unskip
\newblock
\APACjournalVolNumPages{Phys. Rev. D}{109}{7}{074012}.
\newblock
\begin{APACrefDOI} \doi{10.1103/PhysRevD.109.074012} \end{APACrefDOI}
\PrintBackRefs{\CurrentBib}

\bibitem [\protect \citeauthoryear {%
Menezes%
, Benghi~Pinto%
, Avancini%
, Perez~Martinez%
\BCBL {}\ \BBA {} Providencia%
}{%
Menezes%
\ \protect \BOthers {.}}{%
{\protect \APACyear {2009}}%
}]{%
Menezes:2008qt}
\APACinsertmetastar {%
Menezes:2008qt}%
\begin{APACrefauthors}%
Menezes, D\BPBI P.%
, Benghi~Pinto, M.%
, Avancini, S\BPBI S.%
, Perez~Martinez, A.%
\BCBL {}\ \BBA {} Providencia, C.%
\end{APACrefauthors}%
\unskip\
\newblock
\APACrefYearMonthDay{2009}{}{},
\newblock
\unskip
\newblock
\APACjournalVolNumPages{Phys. Rev. C}{79}{}{035807}.
\newblock
\begin{APACrefDOI} \doi{10.1103/PhysRevC.79.035807} \end{APACrefDOI}
\PrintBackRefs{\CurrentBib}

\bibitem [\protect \citeauthoryear {%
Menezes%
, Pinto%
\BCBL {}\ \BBA {} Provid\^encia%
}{%
Menezes%
\ \protect \BOthers {.}}{%
{\protect \APACyear {2015}}%
}]{%
Menezes:2015fla}
\APACinsertmetastar {%
Menezes:2015fla}%
\begin{APACrefauthors}%
Menezes, D\BPBI P.%
, Pinto, M\BPBI B.%
\BCBL {}\ \BBA {} Provid\^encia, C.%
\end{APACrefauthors}%
\unskip\
\newblock
\APACrefYearMonthDay{2015}{}{},
\newblock
\unskip
\newblock
\APACjournalVolNumPages{Phys. Rev. C}{91}{6}{065205}.
\newblock
\begin{APACrefDOI} \doi{10.1103/PhysRevC.91.065205} \end{APACrefDOI}
\PrintBackRefs{\CurrentBib}

\bibitem [\protect \citeauthoryear {%
Strickland%
, Dexheimer%
\BCBL {}\ \BBA {} Menezes%
}{%
Strickland%
\ \protect \BOthers {.}}{%
{\protect \APACyear {2012}}%
}]{%
Strickland:2012vu}
\APACinsertmetastar {%
Strickland:2012vu}%
\begin{APACrefauthors}%
Strickland, M.%
, Dexheimer, V.%
\BCBL {}\ \BBA {} Menezes, D\BPBI P.%
\end{APACrefauthors}%
\unskip\
\newblock
\APACrefYearMonthDay{2012}{}{},
\newblock
\unskip
\newblock
\APACjournalVolNumPages{Phys. Rev. D}{86}{}{125032}.
\newblock
\begin{APACrefDOI} \doi{10.1103/PhysRevD.86.125032} \end{APACrefDOI}
\PrintBackRefs{\CurrentBib}

\end{thebibliography}

\end{document}